\documentclass[reprint, amsmath, amssymb, aps]{revtex4-2}
\usepackage{graphicx}
\usepackage{dcolumn}
\usepackage{bm}
\usepackage{hyperref}
\usepackage{multirow}
\usepackage{amsmath}
\hypersetup{hypertex=true,
            colorlinks=true,
            linkcolor=blue,
            anchorcolor=blue,
            citecolor=blue}

\begin{document}
\preprint{APS/123-QED}

\title{Global prediction of nuclear charge density distributions using deep neural network}
\author{Tian Shuai Shang}
 \affiliation{College of Physics, Jilin University, Changchun 130012, China}
\author{Hui Hui Xie}
 \affiliation{College of Physics, Jilin University, Changchun 130012, China}
\author{Jian Li}\email{jianli@jlu.edu.cn}
\affiliation{College of Physics, Jilin University, Changchun 130012, China}
\author{Haozhao Liang}
\affiliation{Department of Physics, Graduate School of Science,	The University of Tokyo, Tokyo 113-0033, Japan}
\affiliation{RIKEN Interdisciplinary Theoretical and Mathematical Sciences Program, Wako 351-0198, Japan}

\date{\today}

\begin{abstract}

    A deep neural network (DNN) has been developed to generate the distributions of nuclear charge density, utilizing the training data from the relativistic density functional theory and incorporating available experimental charge radii of 1014 nuclei into the loss function. The DNN achieved a root-mean-square (rms) deviation of 0.0193 fm for charge radii on its validation set. Furthermore, the DNN can improve the description in both the tail and central regions of the charge density, enhancing agreement with experimental findings. The model's predictive capability has been further validated by its agreement with recent experimental data on charge radii. Finally, this refined model is employed to predict the charge density distributions in a wider range of nuclide chart, and the parameterized charge densities, charge radii, and higher-order moments of charge density distributions are given, providing a robust reference for future experimental investigations. 
\end{abstract}

\maketitle

\section{\label{sec1}introduction}

    \begin{figure*}[htbp]  
        \centering  
        \includegraphics[width=0.9\textwidth]{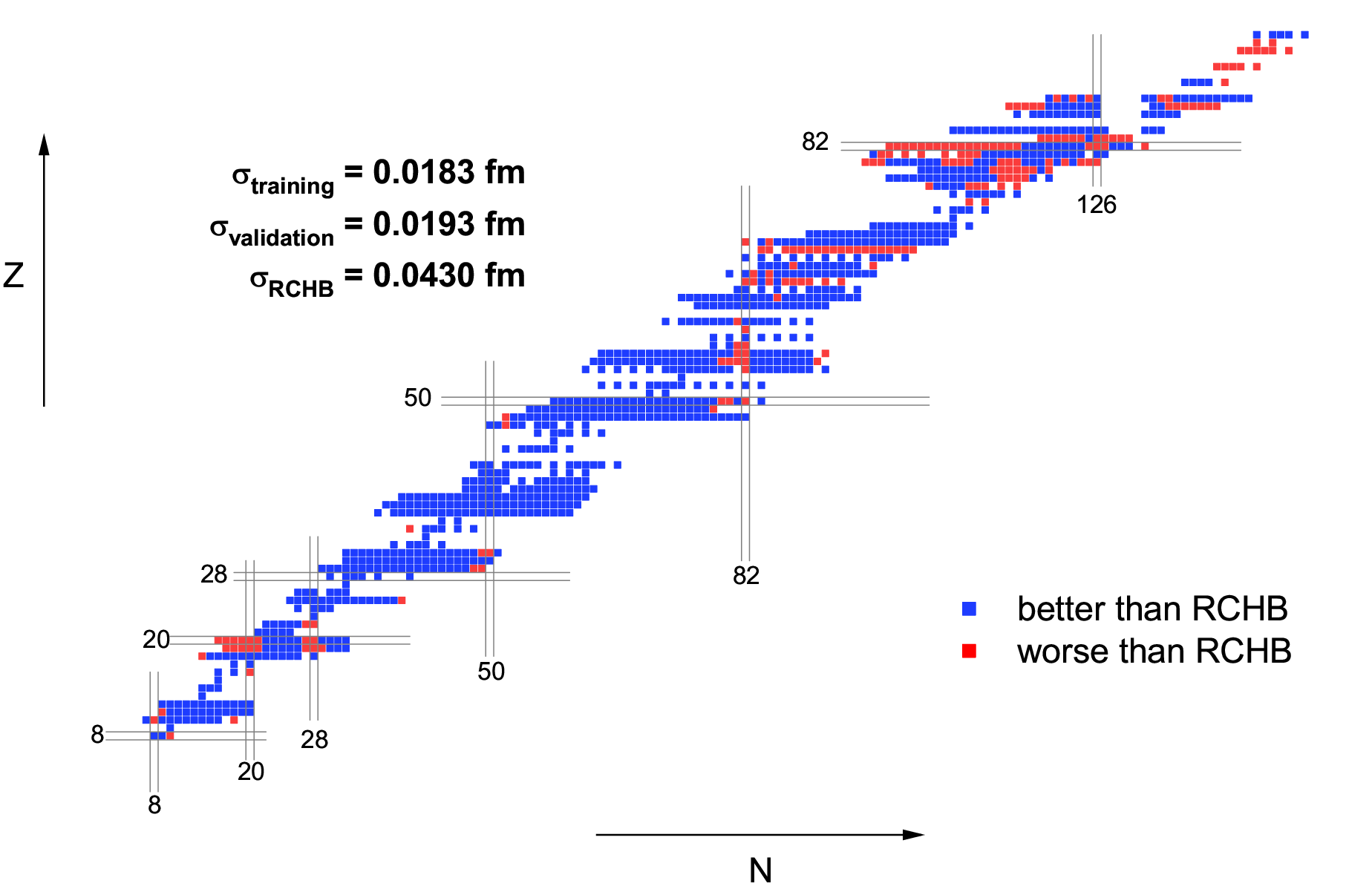}
        \caption{Comparison of charge radii obtained by the deep neural network (DNN) and the relativistic continuum Hartree-Bogoliubov (RCHB) theory on the nuclide chart. The blue (red) squares indicate that the charge radii obtained by DNN are closer to (further deviate from) the experimental values compared with those by the RCHB theory. The mean square error of charge radii by DNN is $0.0183\ \mathrm{fm}$ on the training set and $0.0193\ \mathrm{fm}$ on the validation set, while that of the RCHB method is $0.0430\ \mathrm{fm}$.}\label{figure1}
    \end{figure*}

    Charge density distribution is a fundamental property of atomic nuclei. It not only reflects the abundant nuclear structure information, such as shell structure, shape coexistence, and shape transition but also is an important reference for validating, developing, and perfecting nuclear structure models \cite{Bender2003RMP, Chadwick1932Nature, Wood1992PR, Li2016NST}. 
    The nuclear charge density can also constrain the key parameters in the equation of state of nuclear matter and provide important inputs for nuclear astrophysics research, where nuclear symmetry energy and its density dependence play a significant role in understanding the astrophysical observations \cite{Horowitz2019AP,Chen2019CPC,Arnould2020PPNP}. 
    In atomic physics, the spectroscopic properties of atoms affected by electron-nucleus hyperfine interactions are sensitively dependent on the charge densities of nucleus, so reliable nuclear charge densities are equally critical to understand atomic structure and precise spectroscopic properties \cite{Visscher1997ADNDT,Dirk2000PR,Xie2023PRA}. 
    In addition, the accurate information of nuclear charge density distribution lays a foundation for validating the accuracy of Quantum Electrodynamics under extremely strong electromagnetic field conditions \cite{Gumberidze2005PRL,Sailer2022Nature,Morgner2023Nature}.
    
    High-energy electron elastic scattering is a primary experimental technique for precisely measuring the charge density distribution of nuclei \cite{Ehrenberg1959PR, Meyer1959AP, Kim1992PRC, Wei2021NST}. At present, the experimental distribution is limited to stable nuclei and long-lived unstable nuclei \cite{Vries1987ADNDT}. There are two main methods for fitting experimental data to obtain charge density: model-dependent analysis and model-independent analysis \cite{Jager1974ADNDT, Vries1987ADNDT, Fricke1995ADNDT}. The former include the harmonic-oscillator model, the two-parameter Gaussian model, and the two-parameter Fermi model, which fit fewer parameters and are convenient to use. Because the specific form of the charge density distribution is often unknown, and there is a need to accurately fit experimental data of electron elastic scattering over a large momentum transfer range, model-independent analysis methods are frequently used, such as the Fourier-Bessel (FB) series expansion and sum of Gaussian method \cite{Dreher1974NPA, Sick1974NPA}. The parameters of these empirical models depend on the structural properties of specific nuclei and are usually derived from the experimental data of each nucleus, which lacks the support of microscopic physical mechanisms and is difficult to extrapolate to other nuclei, especially unstable nuclei without experimental data. In view of the demanding description and prediction of experimental data of unstable nuclei and the developing trend of nuclear theoretical models, it is necessary to use microscopic models to study the charge densities of nuclei, develop methods suitable for a wider range of nuclei with more accurate prediction capabilities, and study the micro physical mechanism.

    The most widely used microscopic model to describe nuclear charge properties is density functional theory (DFT), which includes both the non-relativistic \cite{Richter2003PRC, Abdullah2017PJP, Abbas2022BCJ} and relativistic \cite{Ring1996PPNP, Vretenar2005PR, Meng2006PPNP, Meng2013FP, Meng2016BOOK, Li2018FP, Shen2019PPNP,Li2013PRC} DFTs. DFT has become one of the standard theoretical methods for studying nuclear structure \cite{Bender2003RMP}, which can achieve a unified, microscopic, and self-consistent description of almost all nuclei on the nuclide chart without introducing any additional parameters. Nuclear relativistic density functional theory is a relativistic quantum theory based on the effective field theory and DFT to describe nuclear many-body problems. In particular, relativistic DFT takes into account the Lorentz covariance and the time and space components of the corresponding nucleon four-dimensional Lorentz electromagnetic current correspond to the charge density and current, respectively, and can self-consistently and microscopically describe the electromagnetic properties of the atomic nuclei, including the charge density and radius, originating from the electromagnetic density and current. Relativistic DFT has received increasing attention in recent years and has been successfully used to study the ground-state and excited-state properties of stable and exotic nuclei \cite{Xia2018ADNDT,Yang2021PRC,Zhang2022ADNDT,Nikolaus1992PRC,Burvenich2004NPA,Zhao2010PRC,Vretenar2005PR,Meng2006PPNP,Meng2016BOOK,Xie2023PLB, Xie2024PRC, Li2013PRC,Shang2024PLB,Geng2023NST}.
    
    At present, the charge density of nuclei based on relativistic DFT is poorly studied, with the main focus on the charge radius. Relativistic DFT not only gives a good global description of charge radius but also has a strong ability to predict the evolution of some isotopic chains \cite{Geng2003PTP,Agbemava2014PRC,Xia2018ADNDT,Perera2021PRC}. However, the accuracy of the description of charge radius by DFT needs to be further improved compared with the methods that take into account the Garvey-Kelson local relations \cite{Piekarewicz2010EPJA}, and the obtained accuracy varies considerably with different nuclei. Meanwhile, the accuracy of the resulting nuclear charge density distribution still faces challenges, which further limits the application of its charge density. 
    
    Employing neural network techniques can further enhance the accuracy of results from microscopic models based on existing data. Neural networks have been validated as universal data approximators, with deep neural networks (DNN) demonstrating exceptional data processing capabilities. As early as the 1990s, machine learning, and neural networks began to be applied to the modeling of observational data in nuclear physics and have been widely adopted across various fields \cite{Gazula1992NPA, Gernoth1993PLB}. Summarizing these studies, it is concluded that the application of machine learning techniques in the field of nuclear physics requires not only meticulous consideration of model construction but also the integration of physical information into the networks \cite{Boehnlein2022RMP,He2023SCPMA,He2022SSPMA}. This approach ensures that these models not only serve as excellent data fitters but also generate accurate extrapolative data. In the realm of nuclear charge distribution, machine learning methods have also shown impressive capabilities. In studies of charge radii, artificial neural networks \cite{Akkoyun2013JPGNPP, Wu2020PRC}, Bayesian neural networks \cite{Utama2016JPGNPP, Dong2022PRC, Dong2023PLB}, convolutional neural networks \cite{Su2023Sym, Cao2023NST}, and kernel ridge regression \cite{Ma2022CPC, Tang2024NST} methods have all achieved considerable accuracy. Moreover, existing research confirms the promising application prospects of machine learning methods in density distribution. Whether by directly learning density distribution values \cite{Yang2021PLB, Yang2023PRC}, or by fitting empirical model parameters \cite{Shang2022NST} or density functionals \cite{Yang2023PLB} to derive density distributions, machine learning has demonstrated its ability. However, there are challenges, including the lack of globally applicable charge density models and a unified description of charge density and charge radius.
    
    In this paper, we construct a DNN model with four hidden layers, which can build a complex mapping relationship between inputs and outputs through multiple combinations of simple nonlinear functions, and use this to train the nuclear charge density distribution derived from the relativistic continuum Hartree-Bogoliubov (RCHB) theory \cite{Meng1996PRL, Meng1998NPA, Vretenar2005PR, Xia2018ADNDT}, which is given in the form of FB coefficients. Finally, the information of the experimental charge radii is incorporated into the network to constrain the charge density, which makes it possible to improve the accuracy of the final results.
    
    The basic formulae of FB expansion and DNN methods are presented in Sec. \ref{sec2}, the results are shown in Sec. \ref{sec3}, and the summary and perspectives are presented in Sec. \ref{sec4}.

\section{\label{sec2}THEORETICAL FRAMEWORK}
    
    \subsection{The Fourier-Bessel Analysis}
    The Fourier-Bessel series expansion was introduced by Dreher et al \cite{Dreher1974NPA}. For practical reasons, the nuclear charge density $\rho_c(r)$ is assumed to be zero beyond a certain cutoff radius $R$. The first $N$ $\left(=R q_{\max } / \pi\right)$ coefficients $a_\nu\ (\nu=1, 2, ..., N)$ of this series expansion are determined directly from the experimental data. The FB expansion of density distribution  with the spherical symmetry imposed reads
    \begin{equation}\label{eq-FB}
        \rho_c(r)=\left\{\begin{array}{lll}
        \sum\limits_{\nu=1}^N a_\nu j_0(\frac{\nu \pi r}{R}) & \text { for } \quad r \leqslant R, \\
        \\
        0                                       & \text { for } \quad r > R,
                       \end{array}\right.
    \end{equation}
    where $j_0(x)=(\sin x)/x$ denotes the spherical Bessel function of order zero. The normalization gives
    \begin{equation}\label{eq-FB-norm}
	   4\pi\int_0^\infty\rho_c(r)r^2dr=4\pi\sum_{\nu=1}^N\frac{(-1)^{\nu+1}a_\nu R^3}{\left(\nu\pi\right)^2}=Z, 
    \end{equation}
    where $Z$ is the proton number. The coefficient $a_\nu$ can be directly determined by the charge form factor: 
    \begin{equation}\label{eq-anu-Fc}	
        a_\nu=\frac{q_\nu^2}{2\pi R}F_c(q_\nu)~~\text{with}~~q_\nu=\frac{\nu\pi}{R}.
    \end{equation}
    The charge form factor $F_c(q)$ can be regarded as the representation of charge density distribution in momentum space, and it is given by a FB transformation of charge density,
    \begin{equation}\label{eq-Fc-rhoc}
	   F_c(q)=\frac{4\pi}{Z}\int_0^\infty\rho_c(r)j_0(qr)r^2dr.
    \end{equation}
    Combining Eq. (\ref{eq-anu-Fc}) with Eq. (\ref{eq-Fc-rhoc}), the FB coefficients can be determined directly when the charge density is given. By expanding $\rho_c(r)$ into FB series with finite terms using Eq. (\ref{eq-FB}), one can calculate the $n$-th moment $R_n$ of the charge density distribution as
    \begin{equation}\label{eq-Rn}
        R_n \equiv\left\langle r_c^n\right\rangle=\frac{4 \pi}{Z} \int_0^{R} \rho_c(r) r^2 r^n d r.
    \end{equation}
    In particular, $R_c$ is used in this paper to represent the square root of the second moment, i.e., the charge radius
    \begin{equation}\label{eq-Rc}
        R_c \equiv \sqrt{\left\langle r_c^2\right\rangle}.
    \end{equation}

    To get the dataset of nuclear charge density distribution, systematic spherical calculations over the nuclide chart in the framework of RCHB theory with PC-PK1 \cite{Zhao2010PRC} is performed. As one of the most successful relativistic energy density functionals, PC-PK1, which is fitted to the binding energies, charge radii, and empirical pairing gaps of 60 selected spherical nuclei, has been used successfully in describing not only nuclear ground-state properties \cite{Zhang2014FP,Lu2015PRC,Zhao2012PRC} but also various excited-state properties \cite{Yao2013PLB,Yao2014PRC,Wu2014PRC,Li2012PLB,Fu2013PRC,Li2013PLB,Xiang2013PRC,Wang2015JPGNPP}. In particular, PC-PK1 provides a good description for the isospin dependence of binding energy along either the isotopic or isotonic chain, which makes it more reliable for describing exotic nuclei \cite{Zhao2010PRC,Zhao2012PRC}. After taking into account the intrinsic nucleon contributions and nucleon spin-orbit contribution, the relativistic nuclear charge density $\rho_c$ can be self-consistently constructed \cite{Xie2023ARXIV}.

    \begin{table}
    \renewcommand{\arraystretch}{1.3}
    \footnotesize
    \caption{The hyperparameter set of DNN.}\label{tb-hp-prmt}
    \begin{tabular}{llll}
        \hline 
        \hline 
         \makebox[0.12\textwidth][l]{The $i$-th layer} & 
         \makebox[0.13\textwidth][l]{Name} & 
         \makebox[0.1\textwidth][l]{\begin{tabular}{l} Number of \\ neurons \end{tabular}} & 
         \makebox[0.12\textwidth][l]{\begin{tabular}{l} Activation \\ functions \end{tabular}} \\
        \hline 
        0 & Input Layer & 2 & \\
        1 & Dense Layer & 20 & Tanh \\
        2 & Dense Layer & 100 & Tanh \\
        3 & Dense Layer & 100 & Tanh \\
        4 & Dense Layer & 20 & Tanh \\
        5 & Output Layer & 17 &  \\
        \hline 
        \multicolumn{2}{l}{Other hyperparameters} & \multicolumn{2}{l}{Values and properties} \\
        \hline 
        \multicolumn{2}{l}{Normalization Factor} & \multicolumn{2}{l}{$14.85$} \\
        \multicolumn{2}{l}{Batch Size} & \multicolumn{2}{l}{16} \\
        \multicolumn{2}{l}{Objective (Target) Function} & \multicolumn{2}{l}{Loss$_1$, Loss$_2$} \\
        \multicolumn{2}{l}{Optimizer} & \multicolumn{2}{l}{Adam} \\
        \multicolumn{2}{l}{$\lambda$ in Eq.~(\ref{eq-Loss2})} & \multicolumn{2}{l}{$0.7$} \\
        \multicolumn{2}{l}{Learning Rate} & \multicolumn{2}{l}{$l r=5 \times 10^{-3}$ for Loss$_1$} \\
        \multicolumn{2}{l}{ } & \multicolumn{2}{l}{$l r=5 \times 10^{-4}$ for Loss$_2$} \\
        \hline
        \hline
    \end{tabular}
    \end{table}

    Using RCHB to calculate nuclear charge density distributions and calculating the corresponding FB coefficients from Eqs. (\ref{eq-FB}) and (\ref{eq-FB-norm}), the dataset that can be learned by the neural network will be obtained. Meanwhile, the experimental charge radius is introduced into the loss function to embed the experimental information, and the charge density is constrained to make it more consistent with the experimental data, so as to obtain a more reliable global description of charge density distributions. To do this, deep neural networks are reliable and efficient.
    
    \subsection{The Deep Neural Network Approach}
    In the present study, we build a six-layer fully connected neural network, which includes an input layer, an output layer, and four hidden layers. Each layer contains multiple neurons, which receive incoming information and pass it to the next layer through an activation function. The information ultimately reaches the output layer. The network calculates the target function (i.e., the loss function) to obtain the error between the target and the actual network output and propagates the error gradually up the layer through the backpropagation algorithm to update the weight and bias parameters between layers. After several iterations, the final network output is close to the target value and tends to be stable, and then the network training can be considered complete. The carefully modulated hyperparameter set is listed in Table \ref{tb-hp-prmt}.

    The charge radii of about 1,000 nuclei have been measured \cite{Angeli2013ADNDT, Li2021ADNDT}. Among them, 67 nuclei are given by FB analysis, with a maximum of 17 coefficients \cite{Vries1987ADNDT}. Therefore, the number of output neurons of DNN is selected as 17, standing for each of FB coefficients. In order to realize the return of experimental charge radius information to charge density, the training of the neural network is divided into two steps in this work. Firstly, the network learns the FB coefficients of charge density distribution derived from RCHB, and the loss function used is the mean square error (MSE) function, which reads
    \begin{equation} \label{eq-Loss1}
        \operatorname{Loss}_1\left(\boldsymbol{y}_{\mathrm{tar}}, \boldsymbol{y}_{\mathrm{pre}}\right)=\frac{1}{N_{\mathrm{s}}} \sum_{i=1}^{N_{\mathrm{s}}}\left(\boldsymbol{y}_{\mathrm{tar}}-\boldsymbol{y}_{\mathrm{pre}}\right)^2,
    \end{equation}
    where $\boldsymbol{y}_{\mathrm{tar}}$ and $\boldsymbol{y}_{\mathrm{pre}}$ represent the target value (the result of RCHB) and the network output value, respectively. They are all vectors of 17 FB coefficients. ${N_{\mathrm{s}}}$ is the batch-size hyperparameter of the network, which does not have a significant impact during this study and is selected as 16. Through this step, DNN learns the charge density information of RCHB. The second step is to consider the experimental charge radii and incorporate this information into the loss function on the basis of the existing neural network that has been trained by RCHB data. The expression of the loss function is given as
    \begin{equation}
        \begin{aligned}
        \operatorname{Loss}_2\left(\boldsymbol{y}_{\mathrm{tar}}, \boldsymbol{y}_{\mathrm{pre}}\right) & = (1 - \lambda)\frac{1}{N_{\mathrm{s}}} \sum_{i=1}^{N_{\mathrm{s}}}\left(\boldsymbol{y}_{\mathrm{tar}}-\boldsymbol{y}_{\mathrm{pre}}\right)^2 \\
          &+ \lambda\frac{1}{N_{\mathrm{s}}} \sum_{i=1}^{N_{\mathrm{s}}}\left(R_c^{\mathrm{tar}}-R_c^{\mathrm{pre}}\right)^2,
        \end{aligned} \label{eq-Loss2}
    \end{equation}
    where $\lambda$ is the weight hyperparameter, and $R_c^{\mathrm{tar}}$ is the constraint, that is, the experimental charge radii. The DNN's charge radii $R_c^{\mathrm{pre}}$ can be obtained by combining Eqs. (\ref{eq-FB}) and (\ref{eq-Rn}). The value of $\lambda$ and the learning rates of the two steps can be found in Table \ref{tb-hp-prmt}. Through the above two steps of training, the outputs of DNN are expected to be close to the experimental charge density distribution.

    The whole training process is completed under the PyTorch framework. Each training step converges well within 10,000 epochs, which typically takes around an hour of GPU time for training (performed on the NVIDIA GeForce RTX 3070 Ti), while the predictions take just a few milliseconds.

\section{\label{sec3}Results and discussion}

    \begin{figure}[htbp]
        \centering
        \includegraphics[width=0.5\textwidth]{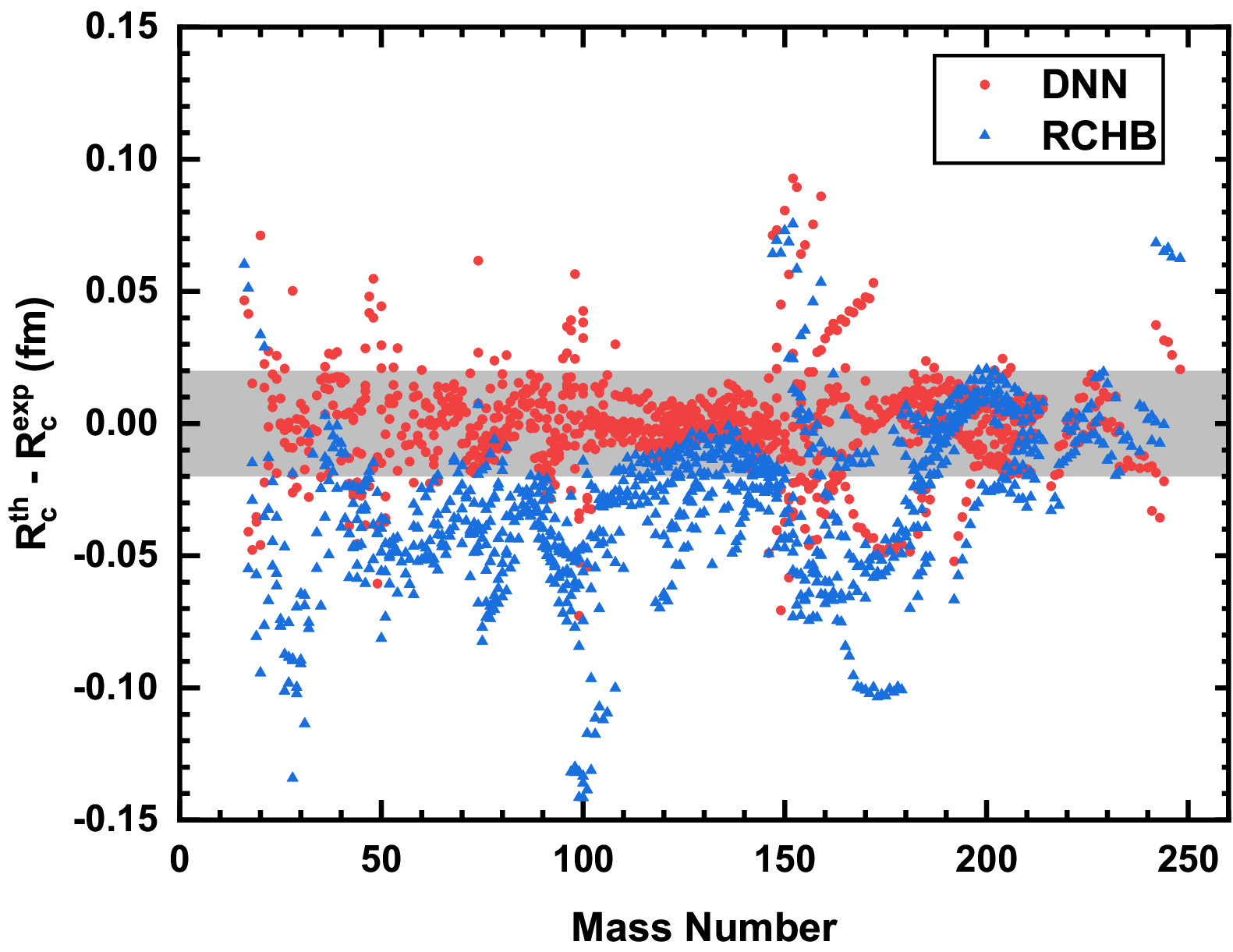}
        \caption{Differences in charge radii from DNN and RCHB relative to experimental values. The red dots indicate deviations for DNN, and blue triangles for RCHB. The gray shaded area represents a range of $\pm 0.02\ \mathrm{fm}$.}
        \label{figure2}
    \end{figure}

    \begin{table}
    \renewcommand{\arraystretch}{1.2}
    \footnotesize
    \caption{Comparison of charge radii of the Ni, Pd, Hg, and Bi isotopes obtained by DNN with the corresponding RCHB and experimental values. The corresponding experimental data are from the references mentioned in the last column.}\label{tb-pred}
    \begin{tabular}{lllllll}
        \hline
        \hline
        \rule{0pt}{11pt} 
        \makebox[0.04\textwidth][l]{$Z$} & 
        \makebox[0.04\textwidth][l]{$A$} & 
         \makebox[0.04\textwidth][l]{$N$} & 
         \makebox[0.08\textwidth][l]{\begin{tabular}{l} $R_c^{\text {DNN}}$ \\ $(\mathrm{fm})$ \end{tabular}} & 
         \makebox[0.08\textwidth][l]{\begin{tabular}{l} $R_c^{\text {RCHB}}$ \\ $(\mathrm{fm})$ \end{tabular}} & 
         \makebox[0.08\textwidth][l]{\begin{tabular}{l} $R_c^{\text {Exp.}}$ \\ $(\mathrm{fm})$ \end{tabular}} & 
         \makebox[0.04\textwidth][l]{Ref.} \\
        \hline
        \multirow{9}*{28} & 54 & 26 & 3.7851 & 3.7220 & 3.7366 & \cite{pineda2021PRL} \\
        ~ & 55 & 27 & 3.7808 & 3.7074 & 3.7252 & \cite{sommer2022PRL} \\
        ~ & 56 & 28 & 3.7785 & 3.6974 & 3.7226 & \cite{sommer2022PRL} \\
        ~ & 59 & 31 & 3.7967 & 3.7423 & 3.782 & \cite{malbrunot2022PRL} \\
        ~ & 63 & 35 & 3.8618 & 3.8058 & 3.842 & \cite{malbrunot2022PRL} \\
        ~ & 65 & 37 & 3.8828 & 3.8365 & 3.856 & \cite{malbrunot2022PRL} \\
        ~ & 66 & 38 & 3.8895 & 3.8513 & 3.870 & \cite{malbrunot2022PRL} \\
        ~ & 67 & 39 & 3.8949 & 3.8703 & 3.873 & \cite{malbrunot2022PRL} \\
        ~ & 70 & 42 & 3.9099 & 3.8914 & 3.910 & \cite{malbrunot2022PRL} \\
        \hline
        \multirow{8}*{46} & 98 & 52 & 4.4303 & 4.3950 & 4.4192 & \cite{geldhof2022PRL} \\
        ~ & 99 & 53 & 4.3940 & 4.4025 & 4.4316 & \cite{geldhof2022PRL} \\
        ~ & 100 & 54 & 4.4834 & 4.4192 & 4.4532 & \cite{geldhof2022PRL} \\
        ~ & 101 & 55 & 4.4528 & 4.4280 & 4.4646 & \cite{geldhof2022PRL} \\
        ~ & 112 & 66 & 4.5881 & 4.5422 & 4.5957 & \cite{geldhof2022PRL} \\
        ~ & 114 & 68 & 4.6033 & 4.5586 & 4.6094 & \cite{geldhof2022PRL} \\
        ~ & 116 & 70 & 4.6166 & 4.5132 & 4.6189 & \cite{geldhof2022PRL} \\
        ~ & 118 & 72 & 4.6279 & 4.5872 & 4.6268 & \cite{geldhof2022PRL} \\
        \hline
        \multirow{2}*{80} & 207 & 127 & 5.4834 & 5.4952 & 5.4923 & \cite{day2021PRL} \\
        ~ & 208 & 128 & 5.4917 & 5.5100 & 5.5033 & \cite{day2021PRL} \\
        \hline
        \multirow{4}*{83} & 187 & 104 & 5.4253 & 5.4186 & 5.4345 & \cite{barzakh2021PRL} \\
        ~ & 188 & 105 & 5.4330 & 5.4250 & 5.4907 & \cite{barzakh2021PRL} \\
        ~ & 189 & 106 & 5.4400 & 5.4279 & 5.4428 & \cite{barzakh2021PRL} \\
        ~ & 191 & 108 & 5.4519 & 5.4381 & 5.4473 & \cite{barzakh2021PRL} \\
        \hline
        $\sigma$  & & &  0.0273  &  0.0374 \\
        \hline
        \hline
    \end{tabular}
    \end{table}

    \begin{figure*}[htbp]
        \centering
        \includegraphics[width=0.9\textwidth]{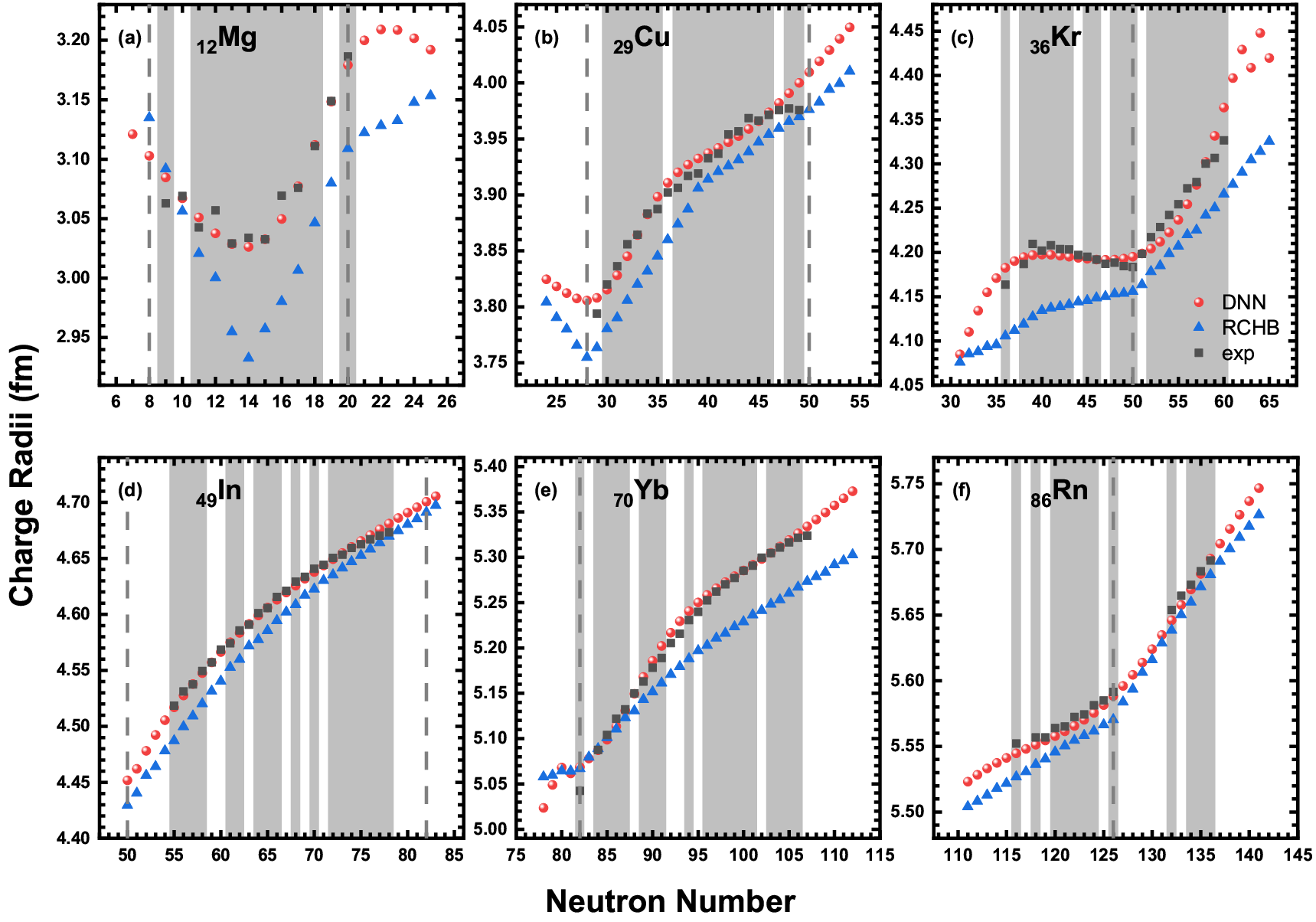}
        \caption{Charge radii predicted by DNN for the $\mathrm{Mg}, \mathrm{Cu}, \mathrm{Kr}, \mathrm{In}, \mathrm{Yb}$, and $\mathrm{Rn}$ isotopes, where the training regions are indicated by shadows and the magic numbers of neutron are indicated by the vertical dashed lines. The RCHB results and the corresponding experimental data are also shown for comparison.}
        \label{figure3}
    \end{figure*}
    
    To assess the impact of information of experimental charge radii on DNN, a total of 1014 measured nuclei \cite{Angeli2013ADNDT, Li2021ADNDT} are used for the dataset. The theoretical FB coefficients of these nuclei are derived from the RCHB calculations. The dataset are randomly split into the training and validation sets in an 8:2 ratio, and this dataset split is fixed for the subsequent training. For the training set, the DNN initially learns the theoretical FB coefficients, and incorporates the experimental charge radii into the loss function for the second step, as mentioned in Sec. \ref{sec2}.

    \begin{figure*}[htbp]
        \centering
        \includegraphics[width=0.9\textwidth]{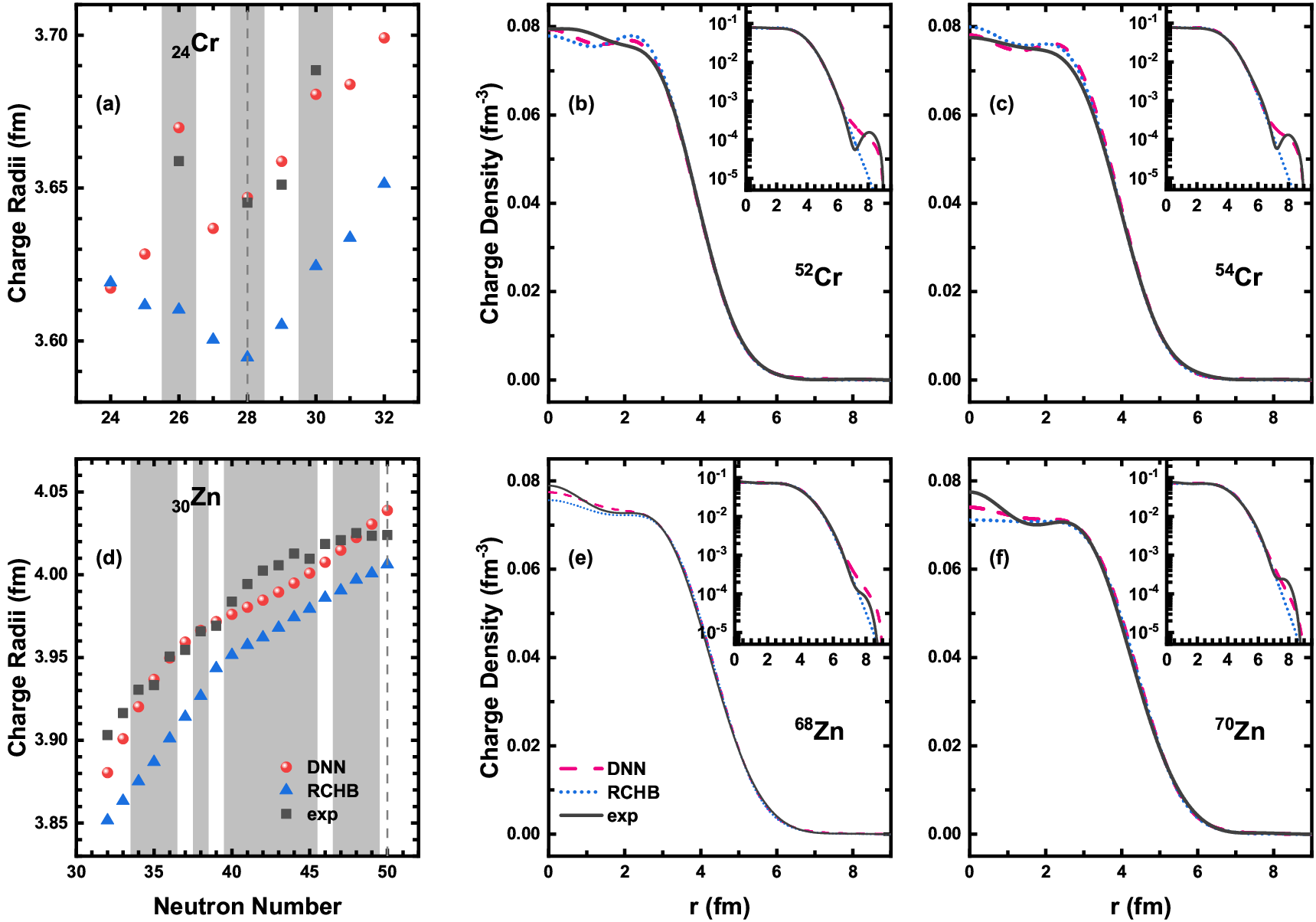}
        \caption{(a) and (d): Charge radii for the $\mathrm{Cr}$ and $\mathrm{Zn}$ isotopes. The training regions are indicated by shadows and the magic numbers are indicated by the vertical dashed lines. (b), (c), (e), and (f): Charge density distributions of $^{52} \mathrm{Cr}, ^{54} \mathrm{Cr}, ^{68} \mathrm{Zn}$, and $^{70} \mathrm{Zn}$. The RCHB results and the corresponding experimental data are also shown for comparison.}
        \label{figure4}
    \end{figure*}

    \begin{figure*}[htbp]
        \centering
        \includegraphics[width=0.9\textwidth]{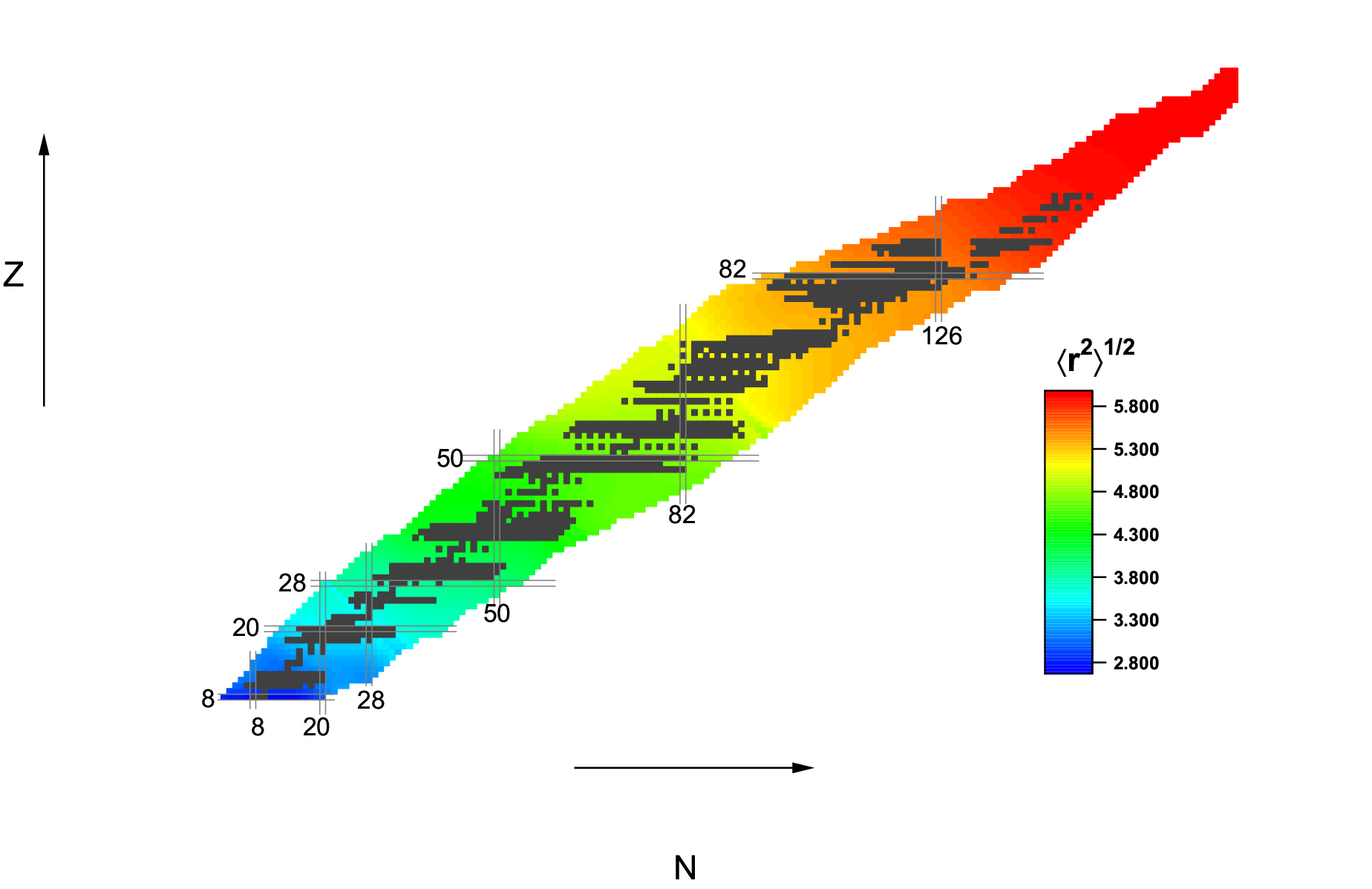}
        \caption{Charge radii predicted by DNN on the nuclide chart. Gray squares denote the nuclei with available experimental charge radii; Others represent charge radii (in $\mathrm{fm}$) calculated by Eq. (\ref{eq-Rn}) with the DNN-inferred charge density distribution.}
        \label{figure5}
    \end{figure*}

    
    The outputs of DNN are the FB coefficients, from which the corresponding charge radius can be calculated using Eqs. (\ref{eq-Rn}) and (\ref{eq-FB}). Figure \ref{figure1} presents the overall comparison of the charge radii calculated by this method and those calculated by $\mathrm{RCHB}$ on the nuclide chart, with the experimental values as the benchmark. If the DNN's result is closer to the experimental charge radius than that of RCHB, it is represented by a blue square; otherwise, it is represented in red. In Fig. \ref{figure1}, it is evident that the DNN shows significant improvements over RCHB in the light-to-medium mass region, especially in the area of $20<Z<65$, where only a few nuclei are slightly worse than RCHB's calculations, mostly distributed at the ends of the isotopic chains. 
    
    It is worth discussing that DNN cannot accurately represent the evolutionary trend of the charge radius of $\mathrm{Ca}$ isotopes: $\mathrm{Ca}$ isotopes exhibit a strong kink structure at $N=28$, a peak between the two closed shells at $N=20$ and $28$, and a rapid increase in charge radius after $N=28$. This evolution may be attributed to complex physical effects, such as the collective effect \cite{Barranco1985PLB, Naito2023PRC}. Since the unique evolution of charge radius in $\mathrm{Ca}$ isotopes is not present in other isotopic chains, it is challenging for neural network methods to replicate such kink structures without overfitting. 
    
    A similar situation occurs around $Z=80$. For $\mathrm{Au}$ isotopes, at $N=107$ and $108$, the experimental charge radius suddenly decreases, indicating a strong deformation exists at $N=107$ \cite{Ma2021PRC}. For $\mathrm{Hg}$ isotopes, there is a noticeable odd-even staggering in the experimental values of charge radius for $N \leq 105$, while the evolution trend becomes smoother for $N>105$. These complex patterns significantly impact the accuracy of DNN. On the one hand, it attempts to learn the corresponding evolutionary trends, but the unconventional numerical fluctuations make it difficult to achieve success; on the other hand, these evolution patterns affect the DNN's judgment on the surrounding isotopic chains, as seen in its unsatisfactory performance in the rich-proton regions of $\mathrm{Tl}$ and $\mathrm{Pb}$ isotopes, where the experiences of $\mathrm{Au}$ and $\mathrm{Hg}$ isotopes lead it to exhibit even more peculiar behavior, but this only yields poorer results.
    

    
    It is noteworthy that the difference of the target function values on the training and validation sets is quite small. As demonstrated in Fig. \ref{figure1}, the root mean square errors (RMSE) on the training and validation sets with respect to the corresponding experimental charge radii are $0.0183\ \mathrm{fm}$ and $0.0193\ \mathrm{fm}$, respectively. This represents a significant improvement over the RMSE of $0.0430\ \mathrm{fm}$ for RCHB. Additionally, the stability of the DNN's results has been validated. Repeatability tests are conducted with random initialization of network parameters, and the fluctuations in the charge radius RMSE do not exceed $5 \%$. The normalization test is also used in the DNN. The error between the proton numbers calculated by Eq. (\ref{eq-FB-norm}) and the actual ones is about 0.2\%, and thus we do not impose the relevant penalties on the network.

    Quantitative data are presented in Fig. \ref{figure2}, where the red (blue) dots represent the deviation of charge radii between the DNN (RCHB) results with respect to the corresponding experimental values. The gray area in the background marks the deviation less than $0.02\ \mathrm{fm}$, which is the error margin of the charge radii of several dozens of nuclei used by RCHB for fitting experimental data \cite{Zhao2010PRC}. As shown in Fig. \ref{figure2}, the corrections made by DNN are significant, especially in the region where the mass number $A<150$. In the range of larger mass numbers, RCHB itself demonstrates a high accuracy, and DNN achieves a comparable or better accuracy.

    DNN has learned the charge density distribution and charge radius information on the entire nuclide chart, which not only makes it perform well from a global perspective but also allows it to show advantages in detail. Figure \ref{figure3} shows the charge radii of the Mg, Cu, Kr, In, Yb, and Rn isotopes obtained from DNN as well as RCHB, and compares with the corresponding experimental values. In several isotopic chains shown in Fig. \ref{figure3}, the variation of charge radius with the number of neutrons, including the evolution trend and values, is in a good agreement with experiment data. In the Kr isotopes, the kink structure at $N=50$ and the previous decline as well as the subsequent sharp rise are successfully evolved by DNN; In the Yb and Rn isotopes, the DNN provides not only a correct evolutionary trend but also a better fit experimental data than RCHB. In addition, to test the extrapolation performance of DNN, several nuclei are extended outward for each isotopic chain in Fig. \ref{figure3}, and most of these results are consistent with the evolutionary trends. Repeatable tests are employed to obtain the error bars for the DNN data, and the network's exceptional stability resulted in error bars smaller than the size of the points in Fig. \ref{figure3} for 100 repeated tests.

    Further, to demonstrate the predictive performance of DNN, we compare the newly measured charge radii on four isotopic chains of $Z=28$, $Z=46$, $Z=80$ and $Z=83$, which are shown in Table \ref{tb-pred}. In the situations where extrapolations by DNN extend several nuclei beyond the training region, the extrapolated charge radii can still be deemed reliable. For example, in the isotopic chains with $Z=46$ and $Z=83$, where extrapolations span more than a dozen nuclei from the training region, most of the predicted outcomes closely match the experimental data. In the isotopic chain of $Z=28$, despite the DNN not delivering sufficiently precise figures, it nevertheless offers a more gradual downward trend within the $N=54-56$ range compared to the RCHB predictions. The last row of Table \ref{tb-pred} shows the rms deviation of the corresponding model with respect to the experimental values. The DNN results are closer to the experimental values than those of RCHB and are optimized by about $0.01\ \mathrm{fm}$.
    
    When focusing on the charge density distribution itself, the DNN also provides satisfactory results. Figure \ref{figure4} displays the outcomes for charge radii and densities in the Cr and Zn isotopes. Given the lack of experimental data on charge density, two nuclei per isotopic chain are selected for detailed presentation. Besides offering improved data on charge radii, the DNN also makes good corrections to the charge density. On the one hand, the DNN optimizes the central density predicted by RCHB, making it more consistent with the experimental distribution; on the other hand, the DNN is capable of replicating the distribution of the experimental density tails, as demonstrated in Figs. \ref{figure4}(b) and (c). While the RCHB density shows an exponential decline, the experimental density exhibits a tail. The DNN, by leveraging information from experimental charge radii, keenly captures this feature and reproduces it accurately.

    Finally, the well-trained deep neural network is utilized to predict nuclear charge density distributions. The prediction range is selected to include nuclei listed in AME2020 with $Z \geq 8$. Figure \ref{figure5} shows the specific distribution of the prediction range. These results are detailed in the supplemental material.

\section{\label{sec4}SUMMARY AND PROSPECTS}

    A deep neural network model is trained to generate nuclear charge density distributions. The training data for the model is based on the distributions generated by the RCHB theory, together with 1014 experimental data of charge radii. The model not only provided more accurate charge radii on a global scale but also corrected the density distribution curves to align more closely with existing experimental outcomes. The predictive capability of the model is validated by comparing it with the recently measured experimental data on charge radii. Finally, the model is used to predict a broader range of charge density distributions, offering a reference for future experiments.

    While the model achieved overall high accuracy, it still failed to replicate specific features such as the kink structure in the charge radii of Ca isotopes and the dramatic decline in charge radii near $Z = 80$. Furthermore, interpreting the results of the neural network remains a challenge. How to better apply machine learning methods to physics research remains an open question that requires further investigation in the future.
    
\begin{acknowledgments}
This work is supported by the Natural Science Foundation of Jilin Province (Grant No. 20220101017JC), the National Natural Science Foundation of China (Grant No. 11675063), the Key Laboratory of Nuclear Data Foundation (JCKY2020201C157), the JSPS Grant-in-Aid for Scientific Research (S) under Grant No.~20H05648,
the RIKEN iTHEMS Program, and the RIKEN Pioneering Project: Evolution of Matter in the Universe.
\end{acknowledgments}

\nocite{*}

\bibliography{apssamp}

\end{document}